\documentclass[prd,aps,preprint,tightenlines,superscriptaddress]{revtex4}
\usepackage{epsfig}
\usepackage{amsmath}
\preprint{DOE/ER/40762-286} \preprint{UM-PP\#03-057}

\begin{document}
\title{Sivers Function in the MIT Bag Model}
\author{Feng Yuan}
\email{fyuan@physics.umd.edu}
\affiliation{Department of Physics,
University of Maryland,
College Park, Maryland 20742}
\affiliation{Institute for Nuclear Theory,
University of Washington,
Box 351550, Seattle, WA 98195}
\date{\today}
\begin{abstract}

The Sivers function, an asymmetric transverse-momentum distribution
of the quarks in a transversely polarized nucleon, is calculated in the MIT 
bag model. The bag quark wave functions contain both $S$-wave and $P$-wave
components, and their interference leads to nonvanishing Sivers
function in the presence of the final state interactions. 
We approximate these interactions through one-gluon exchange. 
An estimate of another transverse momentum dependent
distribution $h_1^\perp$ is also performed in 
the same model. \vspace{10cm}
\end{abstract}
\maketitle
\newcommand{\be}{\begin{equation}}
\newcommand{\ee}{\end{equation}}
\newcommand{\ben}{\[}
\newcommand{\een}{\]}
\newcommand{\beqn}{\begin{eqnarray}}
\newcommand{\eeqn}{\end{eqnarray}}
\newcommand{\Tr}{{\rm Tr} }

The measurements from the HERMES, SMC, and JLAB collaborations show a
remarkably large single-spin asymmetry (SSA) in the semi-inclusive
processes, such as pion production in $\gamma^*p\rightarrow \pi
X$, when the proton is polarized transversely to the direction
of the virtual photon \cite{hermes,smc,jlab}. If the underlying process
is hard, the physical interpretation of such single-spin asymmetry 
can be attributed to either the quark transversity distribution $h_1(x)$
\cite{jaffe-ji} convoluted with the Collins' fragmentation
function $H_\perp(z,k_\perp)$ \cite{collins-frag}, or the Sivers
function $f_{1T}^\perp(x,k_\perp)$ \cite{sivers} convoluted with
the usual fragmentation function $D(z)$, or both.
\cite{mulders-boer,others}. The Sivers function is the asymmetric
distribution of quarks in a transversely polarized proton which
correlates the quark transverse momentum and the proton
polarization vector $\vec{S}_\perp$. The nonvanishing of the
Sivers function has been confirmed recently 
\cite{brodsky,collins,ji1,ji2,mulders2}. The key
ingredient here is that the gauge link in the gauge-invariant
definition of the transverse momentum dependent (TMD) parton
distributions $f(x,k_\perp)$ generates the initial and/or final
state interactions, which results in a phase difference in the
interference between different helicity states of the proton
\cite{brodsky,collins,ji1}. 
There are many other approaches to 
understand SSA in semi-inclusive processes \cite{qiu,review}.
For example, in \cite{Bur02} the SSA is connected to 
the impact parameter dependent parton distribution.

The TMD parton distribution
functions are defined through the quark density matrix \cite{collins-pt},
\begin{eqnarray}
{\cal M} &=& p^+
   \int {\frac{d\xi^-d^2\xi_\perp} {(2\pi)^3}}
        e^{-i(\xi^-k^+-\vec{\xi}_\perp\cdot \vec{k}_\perp)} \nonumber \\
  && \times \langle PS|\overline{\psi}(\xi^-,\xi_\perp)
     {\cal L}^\dagger(\xi^-,\xi_\perp) {\cal L}(0,0_\perp)
\psi(0)| PS\rangle \ ,
\label{density}
\end{eqnarray}
where $S^\mu$ is the polarization vector of the nucleon
normalized to $S_\mu S^\mu=-1$, $p^\mu$ is a light-cone vector such
that $p^-=0$.
The gauge link ${\cal L}^\dagger(\xi^-,\xi_\perp)$ is defined \cite{ji1,ji2},
\begin{eqnarray}
  {\cal L}(\xi^-,{\xi_\perp}) = P \exp\left(-ig\int^\infty_{\xi^-}
        A^+(\eta^-,\xi_\perp) d\eta^-\right) P\exp\left(-ig\int^\infty_{\xi_\perp}
    d\eta_\perp\cdot A_\perp(\eta^-=\infty,
            \eta_\perp)\right)
    \ . \nonumber\\
\label{gl}
\end{eqnarray}
In nonsingular gauges, the second term vanishes.
However, in singular gauges, such as the light-cone gauge,
the second term will contribute.
In the following, we will work in the covariant gauge, and so
we can keep only the first term in the gauge link.

A model calculation of the Sivers function
is needed to demonstrate its existence and its size in the typical
kinematic region. Because the Sivers function contains
the interference between different helicity states of the
proton, the model must contain $S$ wave and $P$ wave
components to generate phase difference, i.e., involving the 
proton wave function component with 
nonzero orbital angular momentum \cite{ji3}.
For example, in the quark-diquark model used in \cite{brodsky},
the proton-quark-scalar coupling contains
$S$ and $P$ wave components, corresponding to
the quark spin parallel and anti-parallel
with the proton spin respectively.
This model has been used to calculate the Sivers function and
other interesting distributions \cite{brodsky,ji1,brodsky-boer,gamberg}.
In this paper, we shall study the Sivers function in the 
MIT bag model \cite{mitbag}.
The bag model contains confine physics and incorporate $SU(6)$ 
spin-flavor structure. More importantly,
the bag model wave function has both $S$ and $P$ wave
components.
Of course, the bag model has a number of
well-known problems, including breaking of chiral symmetry and translational
invariance, etc..
Nonetheless, it approximately generates right the
hadron spectrum \cite{mitbag}; it yields reasonable quark distribution
at low energy scale \cite{bag-pdf};
and it can describe the electromagnetic form factors of the
nucleon \cite{bag-form}.

In the MIT bag model, the quark field
has the following general forms \cite{mitbag},
\begin{equation}
\label{bw}
\Psi_\alpha(\vec{x},t)=\sum\limits_{n>0,\kappa=\pm 1,m=\pm 1/2} N(n\kappa)
\{ b_\alpha(n\kappa m)\psi_{n\kappa jm}(\vec{x},t)+
d_\alpha^\dagger(n\kappa m)\psi_{-n-\kappa jm}(\vec{x},t)\} \ ,
\end{equation}
where $b_\alpha^\dagger$ and $d_\alpha^\dagger$ create quark and anti-quark
excitations in the bag with wave functions,
\begin{equation}
\psi_{n,-1,\frac{1}{2}m}(\vec{x},t)=\frac{1}{\sqrt{4\pi}}
\left (
\begin{array}{l}
i j_0(\frac{\omega_{n,-1}|\vec{x}|}{R_0})\chi_m\\
-\vec{\sigma}\cdot\hat{{x}} j_1(\frac{\omega_{n,-1}|\vec{x}|}{R_0})\chi_m
\end{array}
\right )
e^{-i\omega_{n,-1} t/R_0} \ .
\end{equation}
For the lowest mode, we have $n=1$, $\kappa=-1$, and 
$\omega_{1,-1}\approx 2.04$.
In the above equation, $\vec{\sigma}$ is the $2\times 2$ Pauli matrix,
$\chi_m$ the Pauli spinor, and $R_0$ the bag radius.
$\hat{{x}}$ represents the unit vector in the $\vec{x}$
direction, and $j_i$ are spherical Bessel functions.
Taking the Fourier transformation, we have the momentum space
wave function for the lowest mode,
\begin{equation}
\varphi_{m}(\vec{k})=i\sqrt{4\pi}N R_0^3
\left (
\begin{array}{l}
 t_0(|\vec{k}|)\chi_m\\
\vec{\sigma}\cdot\hat{{k}} t_1(|\vec{k}|)\chi_m
\end{array}
\right ) \ ,
\label{wp}
\end{equation}
where the normalization factor $N$ is,
\begin{equation}
    N=\left(\frac{\omega^3}{2R_0^3(\omega-1)\sin^2\omega}\right)^2 \ .
\end{equation}
The two functions $t_i$, $i=0,1$ are defined as
\begin{equation}
t_i(k)=\int\limits_0^1 u^2 du j_i(ukR_0)j_i(u\omega) \ .
\end{equation}
It can be easily seen from the above equations that the bag model wave 
function Eq.~(\ref{wp}) contains 
both $S$ and $P$ wave components.
The interference between these components will generate a phase difference
under the gauge link contribution.

The Sivers function $f_{1T}^\perp$ represents the asymmetric part of
the transverse momentum
distribution of the quark in a transversely polarized proton,
and can be calculated from the quark density matrix
${\cal M}$ in Eq.~(\ref{density}) through expansion \cite{mulders-boer},
\begin{equation}
{\cal M}=\frac{1}{2M}\left [ f_{1T}^\perp(x,k_\perp)
    \epsilon^{\mu\nu\rho\sigma}\gamma_\mu p_\nu k_\rho S_\sigma+ \dots \right ] \ ,
\end{equation}
where $M$ is the nucleon mass. Inverting the above equation, we obtain,
\begin{eqnarray}
f_{1T}^\perp(x,k_\perp)&=&\frac{M}{2\epsilon^{ij}S^ik^j}\int
    \frac{d\xi^-d\xi_\perp}{(2\pi)^3}
e^{-i(\xi^-k^+-\vec{\xi}_\perp\cdot \vec{k}_\perp)}
\nonumber \\
  &&\times \langle PS_\perp|\overline{\psi}(\xi^-,\xi_\perp)
     {\cal L}^\dagger(\xi^-,{\xi_\perp})\gamma^+{\cal L}(0,0_\perp)
\psi(0)| PS_\perp\rangle \ .
\end{eqnarray}
Since we work in the covariant gauge, only the
first term in the gauge link Eq.~(\ref{gl}) (
the light-cone gauge link) contributes.
Without the gauge link contribution, the Sivers function
vanishes. For example, to the leading order, the above function
has the form,
\begin{equation}
f_{1T}^\perp(x,k_\perp)=\frac{M}{2k^y}\int
\frac{d\xi^-d\xi_\perp}{(2\pi)^3}
e^{-i(\xi^-k^+-\vec{\xi}_\perp\cdot \vec{k}_\perp)}
  \langle PS_x|\overline{\psi}(\xi^-,\xi_\perp)\gamma^+\psi(0)| PS_x\rangle \ ,
\end{equation}
where for convenience,
we have chosen a particular polarization vector $S_x$ representing the
proton is polarized along the $\hat x$ direction.
Inserting the bag model wave functions Eq.~(\ref{bw}), we get the
following results for the leading order contribution to the Sivers
function without the gauge link contribution,
\begin{eqnarray}
f_{1T}^\perp(x,k_\perp)&=&\frac{ME_p}{k^y}\int\frac{d^3k_1}{(2\pi)^3}
    \delta(k_1^+-xP^+)\delta^{(2)}(\vec{k}_{1\perp})-\vec{k}_\perp)\nonumber\\
    &&\times~~ \varphi_m^\dagger(\vec{k}_1)\gamma^0\gamma^+\varphi_{m'}(\vec{k}_1)
    \langle PS_x|a^\dagger_{\alpha m}a_{\alpha m'}| PS_x\rangle \ .
\end{eqnarray}
Plugging in Eq.~(\ref{wp}), we find that
\begin{equation}
\varphi_m^\dagger(\vec{k})\gamma^0\gamma^+\varphi_{m'}(\vec{k})
=\delta_{mm'}\left[t_0^2(k)+2t_0(k)t_1(k)\frac{k_z}{k}+t_1^2(k)\right] \ .
\end{equation}
On the other hand, for a transversely polarized proton state, we have
\begin{equation}
\delta_{mm'}\langle PS_x|a^\dagger_{\alpha m}a_{\alpha m'}| PS_x\rangle =0 \ ,
\end{equation}
as expected.

Expanding the light-cone gauge link to next-to-leading order, we have
\begin{eqnarray}
f_{1T}^{\perp\alpha}(x,k_\perp)&=&\frac{M}{2k^y}\int
\frac{d\xi^-d\xi_\perp}{(2\pi)^3}
e^{-i(\xi^-k^+-\vec{\xi}_\perp\cdot \vec{k}_\perp)}
\nonumber \\
  &&\times
\langle PS_x|\overline{\psi}_{\alpha i}(\xi^-,\xi_\perp)(ig)
\int\limits_{\xi^-}^\infty d\eta^- A^+_a(\eta^-,\xi_\perp)T^a_{ij}
     \gamma^+\psi_{\alpha j}(0)| PS_x\rangle +h.c.\ ,
\end{eqnarray}
where $\alpha$ is the flavor index, $i$ the color index, and $T^a$ the
$SU_c(3)$ Gell-Mann matrix.
$g$ is the gluon coupling with quark field in the MIT bag model.

Inserting the MIT bag model wave functions, we find that
\begin{eqnarray}
f_{1T}^{\perp\alpha}(x,k_\perp)&=&-(ig)^2\frac{ME_P}{k^y}\int
\frac{d^3k_1d^3k_3}{(2\pi)^6}
\frac{d^4q}{(2\pi)^4}\delta(k_1^++q^+-xP^+)
\delta^{(2)}(k_{1\perp}+q_\perp-k_\perp)(2\pi)\delta(q^0)\nonumber\\
&&\times \frac{i}{q^++i\epsilon}\frac{-i}{q^2+i\epsilon}
\sum\limits_{\beta,m_1,m_2,m_3,m_4}T^a_{ij}T^a_{kl}
    \langle PS_x |b_{\alpha m_1}^{i\dagger}b_{\alpha m_2}^{j}
    b_{\beta m_3}^{k\dagger}b_{\beta m_4}^{l} |PS_x\rangle\nonumber\\
&&\times\varphi_{m_1}^\dagger(\vec{k_1})\gamma^0\gamma^+
\varphi_{m_2}(\vec{k})\varphi_{m_3}^\dagger(\vec{k_3})\gamma^0\gamma^+
\varphi_{m_4}(\vec{k_3}-\vec{q})+ h.c. \ ,
\label{f1t}
\end{eqnarray}
where $b_{\alpha m}^i$ is the annihilation operator for a quark with flavor $\alpha$,
helicity $m$, and color index $i$.
In the above derivation, we have used the free gluon propagator
as an approximation.
Actually, in the bag the gauge boson propagate differently as in
the vacuum \cite{bag-dynamics}.
The corresponding diagrams for Eq.~(\ref{f1t}) are shown in Fig.~1.
\begin{figure}[t]
\centerline{\psfig{figure=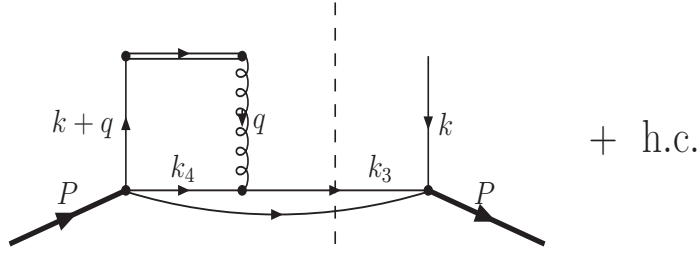,width=9cm,height=3.5cm}}
\vskip 0.2cm
\caption{The leading contribution to the Sivers function $f_{1T}(x,k_\perp)$
in the MIT bag model.}
\end{figure}

Using the identity,
\begin{equation}
\frac{1}{q^+-i\epsilon}-\frac{1}{q^++i\epsilon}=i(2\pi)\delta(q^+)\ ,
\end{equation}
we get
\begin{eqnarray}
f_{1T}^{\perp\alpha}(x,k_\perp)&=&-2g^2\frac{ME_P}{k^y} \int
\frac{d^2q_\perp}{(2\pi)^5}\frac{i}{q^2}
\sum\limits_{\beta,m_1,m_2,m_3,m_4}T^a_{ij}T^a_{kl}
    \langle PS_x |b_{\alpha m_1}^{i\dagger}b_{\alpha m_2}^{j}
    b_{\beta m_3}^{k\dagger}b_{\beta m_4}^{l} |PS_x\rangle\nonumber\\
&&~~\times\varphi_{m_1}^\dagger(\vec{k}-\vec{q}_\perp)\gamma^0\gamma^+
\varphi_{m_2}(\vec{k})
\int \frac{d^3k_3}{(2\pi)^3}
\varphi_{m_3}^\dagger(\vec{k_3})\gamma^0\gamma^+
\varphi_{m_4}(\vec{k_3}-\vec{q}_\perp) \ .
\label{f1t2}
\end{eqnarray}
The $k_3$ integration only depends on $q$, and so we can write this integral
as a function of $q^2$, and define
\begin{equation}
\int\frac{d^3k_3}{(2\pi)^3}\varphi_{m_3}^\dagger(\vec{k_3})\gamma^0\gamma^+
\varphi_{m_4}(\vec{k_3}-\vec{q})\equiv \frac{1}{\sqrt{2}}F(q^2)\delta_{m_3m_4} \ .
\end{equation}
The function $F(q^2)$ is
\begin{eqnarray}
F(q^2)&=&\frac{16\omega^4}{\pi^2 j_0^2(\omega)(\omega -1)}
\frac{1}{M_P^2}\int d^3k_3\left [t_0(|\vec{k}_3|)t_0(|\vec{k}_3'|)+
\frac{k_3^{\prime z}}{|\vec{k}_3'|}t_0(|\vec{k}_3|)t_1(|\vec{k}_3'|)\right.
\nonumber\\
&&\left.+\frac{k_3^{z}}{|\vec{k}_3|} t_1(|\vec{k}_3|)t_0(|\vec{k}_3'|)+\frac{\vec{k}_3\cdot \vec{k}_3'}
{|\vec{k}_3||\vec{k}_3'|}t_1(|\vec{k}_3|)t_1(|\vec{k}_3'|)\right ] \ ,
\end{eqnarray}
where $\vec{k}_3'=\vec{k}_3+\vec{q}$. It is easy to show that
$F(q^2)\rightarrow 1$ as $q^2\rightarrow 0$.

With $F(q^2)$, we can further simplify Eq.~(\ref{f1t2}) as,
\begin{eqnarray}
f_{1T}^{\perp\alpha}(x,k_\perp)&=&-\sqrt{2}g^2C_\alpha^\prime(S_x)\frac{ME_P}{k^y}
\int \frac{d^2q_\perp}{(2\pi)^5}F(q^2)\frac{i}{q^2}
\varphi_{m_1}^\dagger(\vec{k}-\vec{q}_\perp)\gamma^0\gamma^+
\varphi_{m_2}(\vec{k}) \ ,
\end{eqnarray}
and $C_\alpha^\prime$ is
\begin{eqnarray}
C_\alpha^\prime(S_x)&=&\sum\limits_{\beta,m_3,m_4}\delta_{m_3m_4}
T^a_{ij}T^a_{kl}
    \langle PS_x |b_{\alpha m_1}^{i\dagger}b_{\alpha m_2}^{j}
    b_{\beta m_3}^{k\dagger}b_{\beta m_4}^{l} |PS_x\rangle\nonumber\\
&=&\frac{\delta_{m_1+m_2}}{2}C_\alpha \ ,
\end{eqnarray}
where $\delta_{m_1+m_2}$ factor comes from the fact that the proton is
polarized along the $\hat x$ direction, and $C_\alpha$ is defined
as
\begin{equation}
C_\alpha=\sum\limits_{\beta,m_1,m_3}
T^a_{ij}T^a_{kl}
    \langle PS_x |b_{\alpha m_1}^{i\dagger}b_{\alpha -m_1}^{j}
    b_{\beta m_3}^{k\dagger}b_{\beta m_3}^{l} |PS_x\rangle\ .
\label{alphaq}
\end{equation}
Substituting the above results into Eq.~(20), we find 
\begin{equation}
f_{1T}^{\perp\alpha}(x,k_\perp)=-\sqrt{2}g^2C_\alpha\frac{ME_P}{k^y}
\int \frac{d^2q_\perp}{(2\pi)^5}F(q^2)\frac{i}{q^2}
\frac{1}{2}\sum\limits_m\varphi_{m}^\dagger(\vec{k}-\vec{q}_\perp)\gamma^0\gamma^+
\varphi_{-m}(\vec{k}) \ .
\end{equation}
From the bag model wave functions, we obtain
\begin{eqnarray}
\varphi_{m}^\dagger(\vec{k}')\gamma^0\gamma^+
\varphi_{-m}(\vec{k})&=&\frac{4\pi^2R_0^6}{\sqrt{2}} i\chi_m^\dagger
    \left[\frac{(\vec{\sigma}\times \vec{k}')^z}{|\vec{k}'|}t_1(k')t_0(k)
-\frac{(\vec{\sigma}\times \vec{k})^z}{|\vec{k}|}t_0(k')t_1(k)\right.\nonumber\\
    &&\left.+\frac{\vec{k}\times \vec{q}_\perp\cdot\vec{\sigma}}
        {|\vec{k}'||\vec{k}|}t_1(k')t_1(k)\right]\chi_{-m} \ ,
\end{eqnarray}
where $\vec{k}'=\vec{k}-\vec{q}_\perp$.
Because the proton is polarized along the $\hat x$ direction,
in the above equation, only the $\sigma_x$ terms contribute, and
the $q_\perp$ integral will be proportional to $k^y$. And finally, 
we get
\begin{eqnarray}
f_{1T}^{\perp\alpha}(x,k_\perp)&=&-
\frac{16\omega^4g^2C_\alpha}{\pi^2 j_0^2(\omega)(\omega
-1)}\frac{1}{M_P} \left[ I_0(x,k_\perp)-I_1(x,k_\perp)\right]\ .
\label{f1t3}
\end{eqnarray}
The two integrals $I_0$ and $I_1$ are defined as
\begin{eqnarray}
I_0(x,k_\perp)&=&\int \frac{d^2q_\perp}{(2\pi)^2} \frac{F(q_\perp^2)}{q_\perp^2}
\left[\frac{t_1(k')t_0(k)}{k'}-\frac{t_0(k')t_1(k)}{k}\right] \ , \nonumber\\
I_1(x,k_\perp)&=&\left[t_0(k)+\frac{k_z}{k}t_1(k)\right]
\int \frac{d^2q_\perp}{(2\pi)^2} \frac{F(q_\perp^2)}{q_\perp^2}
\frac{t_1(k')}{k'}\frac{\vec{k}_\perp\cdot \vec{q}_\perp}{k_\perp^2} \ ,
\end{eqnarray}
where $k_z=xM_p-\varepsilon$, $k=\sqrt{(xM_P-\varepsilon)^2+k_\perp^2}$ with
$\varepsilon=\omega/R_0$, and
$k'=\sqrt{(xM_P-\varepsilon)^2+(\vec k_\perp-\vec q_\perp)^2}$.
$R_0$ and $M_P$ are bag radius and
proton mass, respectively. In our calculations, we fix the dimensionless
parameter $R_0 M_P=4\omega$.

Working out the matrix elements of Eq.~(\ref{alphaq}) for
the valence quarks, we find
\begin{equation}
    C_u=-\frac{16}{9},~~~~  C_d=\frac{4}{9}\ ,
\end{equation}
for up and down quarks respectively, which means that
the up quark and down quark have opposite
signs for the Sivers function, and differ by a factor of 4.
This is the result of the $SU(6)$ wave function we used for the
proton. For a polarized proton, the polarized up quark distribution 
has a factor of $4/3$ while down quark has $-1/3$.
Phenomenologically,
since $\pi^+$ production is dominated by the up quark fragmentation
and $\pi^0$ is dominated by either the up quark or the down quark
fragmentation,
while $\pi^-$ is dominated by the down quark fragmentation,
the above prediction will lead to larger single spin
asymmetries for $\pi^+$ and $\pi^0$ than that for $\pi^-$
with opposite signs if assuming the Sivers mechanism.
In this estimate, we have neglected the ``unfavored'' fragmentation
contribution to the pion production, which has been shown to
play an important role for $\pi^-$ asymmetry \cite{ma}.
Taking into account the ``unfavored'' ($u$ quark) fragmentation 
contribution which has opposite sign from the ``favored''
($d$ quark) one, we will get even smaller asymmetry for $\pi^-$.
We note that the HERMES collaboration actually
showed much larger asymmetries for $\pi^+$ and $\pi^0$ than
that for $\pi^-$ \cite{hermes}.
On the other hand, concerning the quark distribution
in the neutron, one shall have 4 times 
larger Sivers function for down quark than that for up quark
by isospin symmetry argument from the above results,
and both of them will have different signs compared to the
proton ones. 
That means, with the neutron target, one would have a factor of
2 smaller asymmetry for $\pi^+$ with opposite sign compared to
the asymmetry with the proton target.
It is interesting to note that JLab 
will measure these asymmetries with the proton target,
and the neutron target as well. The comparison of the SSA
between the proton and neutron targets will provide crucial test on
the Sivers mechanism for the SSA.

\begin{figure}[t]
\centerline{\psfig{figure=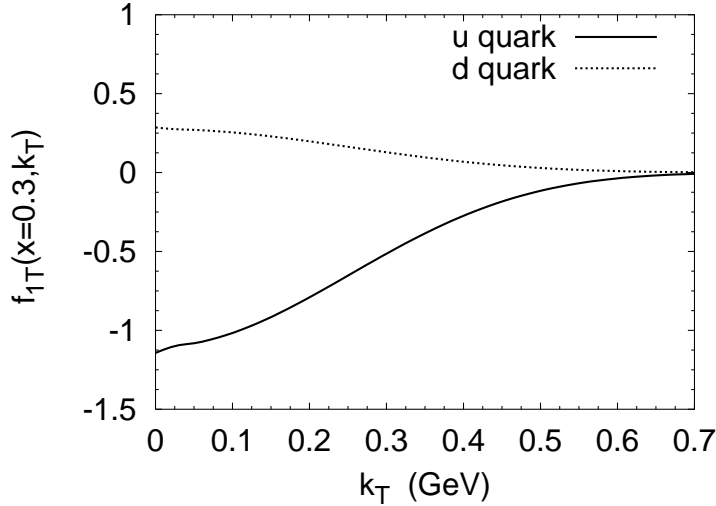,width=10cm}} \vskip 0.2cm
\caption{The Sivers functions $f_{1T}^\perp(x,k_\perp)$ for the
valence quarks at $x=0.3$ as functions of $k_\perp$, where
the quark-gluon coupling $\alpha_s=0.2$.}
\end{figure}

In Fig.~2, we plot the Sivers functions for up and down quarks
as functions of transverse momentum $k_\perp$ at $x=0.3$.
Here the quark-gluon coupling $g$ is treated as a free parameter, and 
we set $\alpha_s =g^2/4\pi=0.2$, which
is smaller than the value used in \cite{mitbag} to
determine the mass splitting of baryons.
Since bag model is not suitable for the calculation of the distribution
at large transverse momentum, here we only show the results for 
the range of $k_\perp$ smaller than $0.7$GeV.

The contribution from the Sivers effect to the SSA in the semi-inclusive 
process can be calculated from the above results divided by the 
unpolarized quark distribution in the same model \cite{bag-pdf}. 
The asymmetry ${\cal P}_y$ is calculated as
\begin{equation}
{\cal P}_y=\frac{k_\perp^x}{M}f_{1T}^\perp(x,k_\perp)/f_1(x,k_\perp) \ ,
\end{equation}
where the polarization of the proton is along the $\hat y$ 
direction. We plot these asymmetries for the valence
quarks as functions of $k_\perp^x$ at $x=0.3$ in Fig.~3(a),
and as functions of $x$ at $k_\perp^x=0.5$ in Fig.~3(b).
These asymmetries are for the quark distributions. To get 
the asymmetry associated with the hadron production in semi-inclusive
processes, we need to convolute the above results with the
fragmentation functions of the hadrons.

\begin{figure}[t]
\centerline{\psfig{figure=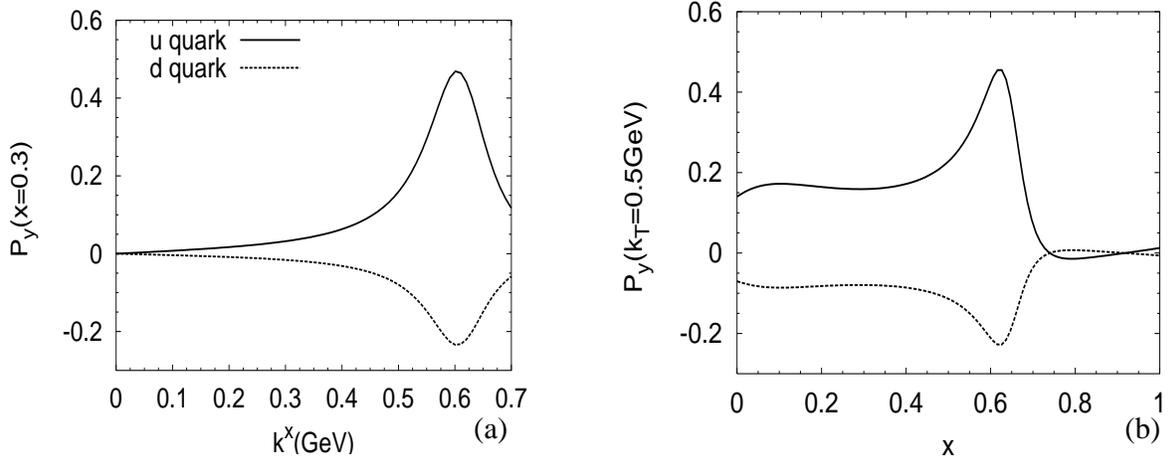,width=16cm}}
\vskip 0.2cm
\caption{The bag model prediction for the asymmetry of the 
quark distribution in a transverse polarized proton as a function
of $k^x$ and $x$, where $\alpha_s=0.2$.}
\end{figure}

It is also interested to study the moments
of the Sivers distribution. For example, one interested moment 
is defined as \cite{evolution}
\begin{equation}
f_{1T}^{\perp(1)}(x)=\int d^2k_\perp
\left(\frac{\vec{k}_\perp^2}{2M^2}\right ) f_{1T}^\perp(x,k_\perp)
\ .
\end{equation}
The numerical results for the above 
functions depending on $x$ are shown in Fig.~4, which can be fit
with the following functional form,
\begin{eqnarray}
\nonumber
f_{1T}^{\perp(1)u}(x)&=&-0.75 x^{1.63}(1-x)^{4.06}\ , \\
f_{1T}^{\perp(1)d}(x)&=& 0.19 x^{1.63}(1-x)^{4.06}\ .
\end{eqnarray}
Since the bag model is not good for small $x$ parton distributions,
we have abandoned the use of small $x$ points in the fit.
As an illustration, we also plot the above fit for the up-quark in 
Fig.~4.

\begin{figure}[t]
\centerline{\psfig{figure=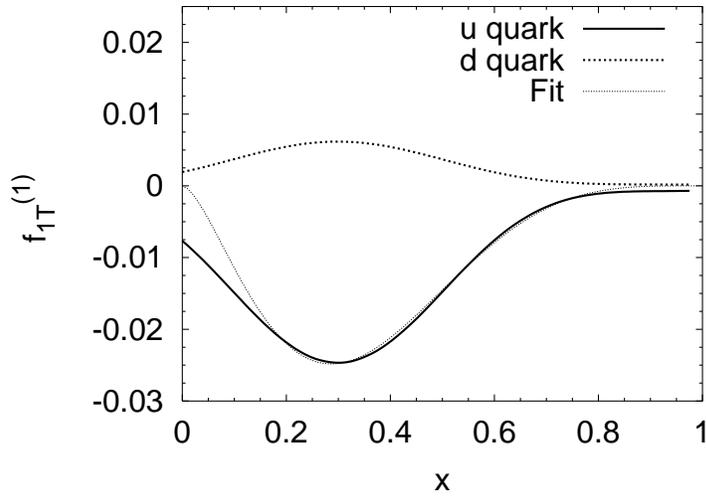,width=10cm}}
\vskip 0.2cm
\caption{The first moments of the Sivers functions $f_{1T}^{(1)}(x)$ for
valence quarks, where $\alpha_s=0.2$.}
\end{figure}

We can repeat the above calculations for another 
TMD parton distribution, $h_1^\perp$,
which represents the correlation between the quark's transverse momentum
and polarization in an unpolarized proton state. It can be
calculated from the expansion of the density matrix,
\begin{equation}
{\cal M}=\frac{1}{2M}\left [h_1^\perp(x,k_\perp)\sigma^{\mu\nu}k_\mu p_\nu
    + \dots\right ] \ .
\end{equation}
Inverting the above equation, we get
\begin{eqnarray}
h_{1}^\perp(x,k_\perp)&=&\frac{M}{2\epsilon^{ij}k^j}\int \frac{d\xi^-d\xi_\perp}{(2\pi)^3}
e^{-i(\xi^-k^+-\vec{\xi}_\perp\cdot \vec{k}_\perp)}
\nonumber \\
  &&\times \langle P|\overline{\psi}(\xi^-,\xi_\perp)
     {\cal L}^\dagger(\xi^-,\xi_\perp)\gamma^+\gamma^i\gamma_5{\cal L}(0,0_\perp)
\psi(0)| P\rangle \ ,
\end{eqnarray}
where the proton is unpolarized.
Without the gauge link contribution, this function vanishes, as the Sivers
function does.
Expanding the gauge link to next-to-leading order, we get
\begin{eqnarray}
h_{1\alpha}^\perp(x,k_\perp)&=&\frac{M}{2\epsilon^{ij}k^j}\int \frac{d\xi^-d\xi_\perp}{(2\pi)^3}
e^{-i(\xi^-k^+-\vec{\xi}_\perp\cdot \vec{k}_\perp)}
\nonumber \\
  &&\times
\langle P|\overline{\psi}_{\alpha i}(\xi^-,\xi_\perp)(ig)
\int\limits_{\xi^-}^\infty d\eta^- A^+_a(\eta^-,\xi_\perp)T^a_{ij}
     \gamma^+\gamma^i\gamma_5\psi_{\alpha j}(0)| P\rangle +h.c. \ .
\end{eqnarray}
where the nucleon is unpolarized.

Using the same method as we did in the calculations of the Sivers function,
we find that
\begin{eqnarray}
h_{1\alpha}^\perp(x,k_\perp)&=&-\sqrt{2}g^2D_\alpha\frac{ME_P}{\epsilon^{ij}k^j}
\int \frac{d^2q_\perp}{(2\pi)^5}F(q^2)\frac{i}{q^2}
\varphi_{m}^\dagger(\vec{k}-\vec{q}_\perp)\gamma^0\gamma^+\gamma^i\gamma_5
\varphi_{m}(\vec{k}) \ ,
\end{eqnarray}
where $D_\alpha$ is defined as
\begin{equation}
    D_\alpha=\sum\limits_{\beta,m_1m_3}T^a_{ij}T^a_{kl}
    \langle P|b_{\alpha m_1}^{i\dagger}b_{\alpha m_1}^{j}
    b_{\beta m_3}^{k\dagger}b_{\beta m_3}^{l} |P\rangle\ .
\end{equation}
And finally, we can write the distribution $h_1^\perp$ in the form of,
\begin{eqnarray}
h_{1\alpha}^\perp(x,k_\perp)&=&-
\frac{16\omega^4g^2D_\alpha}{\pi^2 j_0^2(\omega)(\omega -1)}\frac{1}{M_P}
\left[ I_0(x,k_\perp)-I_1(x,k_\perp)\right]\ ,
\end{eqnarray}
which is the same as Eq.~(\ref{f1t3}) except the color factor.
For the valence quark distributions, we have
\begin{equation}
D_u=-\frac{8}{3},~~~~~D_d=-\frac{4}{3},
\end{equation}
which means that the up and down quarks have the same sign for $h_1^\perp$
distribution, and differ by a factor of two. 
For the unpolarized proton, the up quark distribution is two times larger
than down quark distribution.
This prediction shows that the asymmetries associated with $h_1^\perp$
for $\pi^\pm$ and $\pi^0$ will have the same sign. This is quite different
from the asymmetries associated with the Sivers function $f_{1T}^\perp$
we discussed before.
For the neutron, one has the similar prediction.

As an illustration, we plot in Fig.~5
the distributions $h_1^\perp(x,k_\perp)$ as functions of 
transverse momentum at $x=0.3$.
We can also calculate the moments of the $h_1^\perp$ functions for
up quark and down quark, and fit with the following parameterizations,
\begin{eqnarray}
\nonumber
h_{1}^{\perp(1)u}(x)=-1.13 x^{1.63}(1-x)^{4.06}\ , \\
h_{1}^{\perp(1)d}(x)=-0.56 x^{1.63}(1-x)^{4.06}\ ,
\end{eqnarray}
where the same functional dependence as $f_{1T}^{\perp (1)}$ 
have been observed.

\begin{figure}
\centerline{\psfig{figure=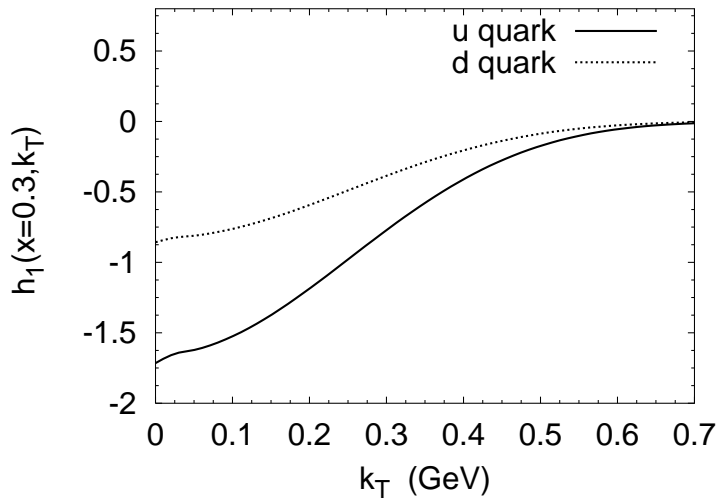,width=10cm}}
\vskip 0.2cm
\caption{ $h_{1}^\perp(x,k_\perp)$ for the
valence quarks at $x=0.3$ as functions of $k_\perp$, where $\alpha_s=0.2$.}
\end{figure}

In conclusion, we have calculated the Sivers
function in the MIT bag model. The gauge link in the 
gauge-invariant definition of the TMD
parton distribution functions plays the crucial role for the
nonvanishing of the Sivers function.
Our calculations show that the up quark Sivers function
is 4 times larger than that of down quark with opposite
signs, consistent with the $SU(6)$ spin-flavor structure of the 
proton. 
These results lead to testable consequence for the single
spin asymmetry associated with the Sivers function in
the semi-inclusive deep inelastic pion productions: 
the asymmetries for $\pi^+$ and $\pi^0$ 
will be larger than that for $\pi^-$, and with 
different signs.
Distribution $h_1^\perp$ has also been calculated in the 
same model, and we found that the up quark and down quark 
have the same sign, which means that the asymmetries 
associated with this distribution
for $\pi^\pm$ and $\pi^0$ will have the same sign.

We end up our paper with a few comments.
First, in our calculations, we have used free gluon propagator
connecting gluon fields inside the bag, which is an approximation
\cite{bag-dynamics}.
Secondly, we have ignored the scale evolution of the Sivers function moments
$f_{1T}^{\perp(1)}(x,Q^2)$ and $h_1^{\perp(1)}(x,Q^2)$ 
\cite{evolution,andrei1}.
Our calculations are performed at the bag scale, which is much lower than
typical hard scattering scales. However, the evolution of these functions
is not clear yet \cite{andrei1,Goe03}, 
and is beyond the scope of the present paper.

The author thanks Andrei Belitsky and Xiangdong Ji for their 
suggestions and useful
comments and critical readings of the manuscript. The author also
thanks Harut Avagyan, Stan Brodsky, Matthias Burkardt,
Xiaodong Jiang, and Mark Strikman for their comments. 
We thank the Department of Energy's Institute for Nuclear Theory at 
the University of Washington for its hospitality during the
program ``Generalized parton distributions and hard exclusive 
processes'' and the Department of Energy for the partial support
during the completion of this paper.
This work was supported by
the U. S. Department of Energy via grants DE-FG02-93ER-40762.

\end{document}